\documentclass[english,superscriptaddress]{revtex4-2}
\usepackage[utf8]{inputenc}
\usepackage{geometry}
\usepackage{amsmath}
\usepackage{amssymb}
\usepackage{graphicx}

\usepackage{times}

\begin{document}
\title{Distributed entanglement generation from asynchronously excited qubits}
\author{Tian-tian Huan}
\affiliation{Institute of Applied Physics and Materials Engineering, University of Macau, Macau, China}
\affiliation{College of Mathematics and Computer Science, Chifeng University, Chifeng 024000, China}
\author{Rigui Zhou}
\affiliation{College of Information Engineering, Shanghai Maritime University, Shanghai 201306, China}
\author{Hou Ian}
\email{houian@um.edu.mo}

\affiliation{Institute of Applied Physics and Materials Engineering, University
of Macau, Macau, China}
\begin{abstract}
The generation of GHZ states calls for simultaneous excitation of
multiple qubits. The peculiarity of such states is reflected in their
nonzero distributed entanglement which is not contained in other entangled
states. We study the optimal way to excite three superconducting qubits
through a common cavity resonator in a circuit such that the generation
of distributed entanglement among them could be obtained at the highest
degree in a time-controllable way. A non-negative measure quantifying
this entanglement is derived as a time function of the quadripartite
system evolution. We find that this measure does not stay static but
obtains the same maximum periodically. When the qubit-resonator couplings
are allowed to vary, its peak value is enhanced monotonically by increasing
the greatest coupling strength to one of the qubits. The period of
its peak to peak revival maximizes when the couplings become inhomogeneous,
thus qubit excitation becoming asynchronous, at a relative ratio of
0.35. The study demonstrates the role  of asynchronous excitations
for time-controlling multi-qubit systems, in particular in extending
entanglement time.
\end{abstract}
\maketitle

\section{Introduction}

GHZ state is used ubiquitously in quantum crytography~\citep{Chen07},
quantum communication~\citep{Gao05}, and metrology~\citep{chin12}.
Its generation among three qubits call for the distributed entanglement
among all qubits~\citep{coffman00} instead of the individual entanglements~\citep{Wootters98}
between one qubit and the other two combined. Generalizing the scenarios
to four or more qubits, the distributed entanglement is formally distinguished
from the individual entanglements in terms of the monogamy relations~\citep{Horodecki09,Osborne06,Cornelio13}.
The former is regarded as the difference between the group entanglement
(one qubit entangles with all the others as a whole) and the sum of
the monogamous entanglements (one qubit entangles with one other qubit).
For an $N$-qubit system, the difference is computed through a metric
called polynomial invariant~\citep{Eltschka12} for its invariance
under local unitary transformations~\citep{Wong01}. It extends the
concept of concurrence~\citep{Mintert051,Mintert052} and generalized
the 3-tangle measure for tripartite systems~\citep{coffman00,Dur00,Regula16,Gartzke18}.
There are also proposals to measure $N$-qubit entanglement through
entanglement formation~\citep{Oliveira14,Bai14}. 

Despite these extensive studies, they are commonly restricted to static
entangled states. The question regarding entanglement dynamics is
less well understood. Using information theoretic measure like distance
function~\citep{Liu17} or concurrence~\citep{Huan15}, the onset
of entanglement accumulation can be found and distinct time behavior
of entanglement evolution can be determined by system parameters such
as coupling strengths for cavity-coupled qubits~\citep{Huan20}.
More recent studies investigate how entanglement can be protected
from the effects of non-Markovian environments through time~\citep{chin12,Nourman16,Zhang10}.
Despite these studies, how entanglement are distributed dynamically
among multiple qubits in a unified system, making distinction between
group entanglement and monogamous entanglement has, to our best knowledge,
not yet been answered.

Here, we study an experimentally accessible circuit quantum electrodynamic
(cQED) system~\citep{wallraff04,fink09} comprising three superconducting
qubits that are commonly coupled to one cavity bus, aiming to clarify
the relationship between the coupling combinations and the temporal
behavior of distributed entanglement. Specifically, we investigate
how multiple qubits undergoing asynchronous excitation by uneven couplings
would facilitate or deteriorate the generation of distributed entanglement.
The deformed algebra technique~\citep{ian12,ian14}, which was used
to study a similar quadripartite system for static entanglement analysis~\citep{Ian16},
is extended to the dynamic analysis. In addition, to accomodate the
quadripartite system, we generalize the 3-tangle definition while
specifying the polynomial invariant definition to introduce a 4-tangle
measure to quantify the distributed entanglement that occurs in our
cQED system. Numerically solving the equations of motion leads to
a cyclic revival pattern of the 4-tangle, which are similar to the
evolution dynamics of various qubit and cavity systems~\citep{ficek06,Dukalski10,Wang14,Huan15}.

More importantly, the individual qubit-cavity coupling strengths are
varied and the resulting maximal 4-tangle and period of revival are
statistically binned to find the optimal coupling strengths. We find
that when the 4-tangle under all circumstances are periodically returned
to zero to ensure the monogamous equality bound, the peak magnitude
of obtainable 4-tangle increases monotonically with the absolute coupling
strength if the couplings among the qubits are set uniform. This observation
provides a positive correlation between monogamy and coupling strength.
Moreover, the length of the period depends nonlinearly on the relative
strength, i.e the ratio of the center qubit coupling to the side qubit
coupling. In particular, statistical analysis shows that the period
maximizes when the relative strength equals $0.35$, showing that
the inhomogeneity of coupling (thus asynchronous excitations to the
qubits) in this case positively affects the generation of entanglement.

Environmental induced decoherences are omitted in our study to simplify
the expressions for the state evolutions and the computation of the
4-tangle. This omission is experimentally justified because typical
superconducting transmon qubits fabricated under current technologies
have $T_{1}$ and $T_{2}$ relaxation times reach the order of 10$\mu$s
and sometimes even $10^{2}\mu$s~\citep{kjaergaard20}. In contrast,
the characteristic times studied here for entanglement generation,
such as the period of revival, is less than 1$\mu$s. Therefore, the
observations from the theoretical study is not affected by the finite
coherence times. In the following, we present the model of the quadripartite
system in Sec.~\ref{sec:model} and define the measure of 4-tangle
in Sec.~\ref{sec:evolution}. The discussion relevant to the periodicity
appearing in the monogamy inequality is presented in Sec.~\ref{sec:periodicity}
before the conclusions are given in Sec.~\ref{sec:Conclusions}.

\section{Quadripartite system\label{sec:model}}

\begin{figure}
\includegraphics[clip,width=8.5cm]{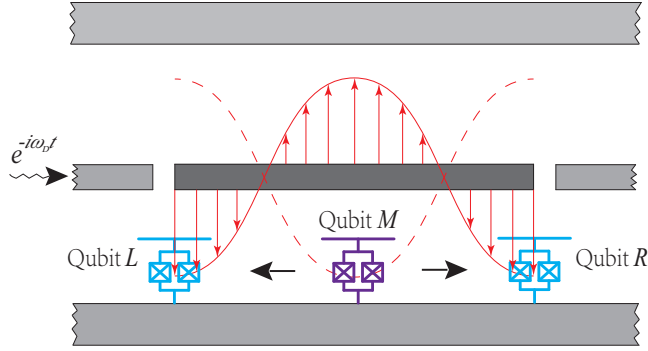}

\caption{Schematic illustration of the quadripartite system: three superconducting
qubits are distributed along a stripline resonator (dark gray rectangle).
Two qubits are located at the antinodes of the cavity field in the
resonator while one in between has a variable location. The resonator
is fed by a microwave driving field from the left along the waveguide.~\label{fig:model}}
\end{figure}

Illustrated in Fig.~\ref{fig:model}, the system comprises a cavity
resonator made of a waveguide stripline and three superconducting
qubits, where we use the indicators $L$ (left), $M$ (middle), and
$R$ (right) to distinguish them. Then, using Pauli matrices for the
two-level qubits of transition frequencies $\{\Omega_{L},\Omega_{M},\Omega_{R}\}$
and a pair of creation and annihilation operators for the cavity field
of frequency $\omega_{c}$, the free energy part of the Hamiltonian
($\hbar=1$) reads $H_{0}=\omega_{c}a^{\dagger}a+\sum_{\nu}\Omega_{\nu}\sigma_{\nu,z}$.
The index $\nu$ ranges over $\{L,M,R\}$. The interaction part corresponds
to the qubit-resonator coupling, in the rotating-wave approximation,
with individual (unequal) coupling strengths $\eta_{L}$, $\eta_{M}$
and $\eta_{R}$, letting the interaction Hamiltonian be $H_{\mathrm{int}}=\sum_{\nu}\eta_{\nu}\left(a\sigma_{\nu,+}+a^{\dagger}\sigma_{\nu,-}\right)$.
In such superconducting circuits, the physical dimension of qubits
(on the scale of $\mu$m) is much less than the inter-qubit spacings
(commensurate with cavity wavelength, on the scale of cm)~\citep{fink09}.
Hence, the direct inter-qubit coupling is neglected.

We consider qubit $L$ and qubit $R$ placed at the edges of the stripline
resonator, i.e. located at antinodes of the cavity field, and thus
always maximally coupled to the cavity field. Qubit $M$ is placed
between the qubits $L$ and $R$ and we allow its location to be variable
such that $\eta_{M}$ be tunable between the maximal coupling attained
by $\eta_{L}$ (and $\eta_{R}$) and the minimal (vanishing) coupling
if it is located at a field node. The cavity field is driven by an
external microwave field with frequency $\omega_{D}$ and a weak driving
amplitude $\varepsilon_{D}$~\citep{Borkje13}, making the external
part of the Hamiltonian be $H_{\mathrm{ext}}=i\varepsilon_{D}\left(a^{\dagger}e^{-i\omega_{D}t}-ae^{i\omega_{D}t}\right)$.

To derive the evolution dynamics of the quadripartite system, we first
diagonalize the closed subsystem consisting of $H_{0}$ and $H_{\mathrm{int}}$.
Under weak driving, only the low-excitation number states $|0\rangle$
and $|1\rangle$ of the cavity mode are considered, giving rise to
16 dressed states, transformable from the tensor product states contributed
by the cavity mode and the qubit eigenstates. Hence, writing the dressed
states as $\left|u_{k}\right\rangle $ with associated eigen-energies
$E_{k}$, we have $(H_{0}+H_{\mathrm{int}})\left|u_{k}\right\rangle =E_{k}\left|u_{k}\right\rangle $,
where the index $k$ ranges over $\{0,...,15\}$. The transformation
between the dressed state and the bare states reads
\begin{equation}
\left|u_{k}\right\rangle =\sum_{\left\langle m\right\rangle }\left[\alpha_{m,0}^{(k)}\left|\phi_{m},0\right\rangle +\alpha_{m,1}^{(k)}\left|\phi_{m},1\right\rangle \right]\label{eq:drs_state}
\end{equation}
where $m$ gives a decimal index converted from the binary combinations
of the qubit states, where the ground state $\left|g\right\rangle $
is designated by 0 and the excited state $\left|e\right\rangle $
by 1. The state of the qubit $L$ (qubit $R$) indicates the most
(least) significant bit, making $m$ range over the integers between
0 and 7. For example, $\left|e_{L},g_{M},g_{R},1\right\rangle =\left|e,g,g,1\right\rangle =\left|\phi_{4},1\right\rangle $.
Also, $\alpha_{m,n}^{(k)}$ indicates the transformation coefficients
for the $k$-th dressed state.

In the space spanned by the basis states of Eq.~(\ref{eq:drs_state}),
the effect of the photonic creation and annihilation are distributed
across all dressed states. Therefore, before we can derive the equation
of motion of the system, we transform the operator $a$ that appears
in $H_{\mathrm{ext}}$ into the dressed basis, i.e.

\begin{eqnarray}
a & = & \mathbb{I}_{L}\otimes\mathbb{I}_{M}\otimes\mathbb{I}_{R}\otimes a\nonumber \\
 & = & \sum_{\left\langle m\right\rangle }|\phi_{m},0\rangle\langle\phi_{m},1|\nonumber \\
 & = & \sum_{j,k}\gamma_{jk}\left|u_{j}\right\rangle \left\langle u_{k}\right|,
\end{eqnarray}
where $\gamma_{jk}=\left\langle u_{j}\right|a\left|u_{k}\right\rangle =\sum_{\left\langle m\right\rangle }\alpha_{m,0}^{(j)*}\alpha_{m,1}^{(k)}$.
Consequently, the total Hamiltonian is written as
\begin{align}
H & =\sum_{k}E_{k}|u_{k}\rangle\langle u_{k}|-i\varepsilon_{D}\sum_{k,j}\left[e^{i\omega_{D}t}\gamma_{kj}|u_{k}\rangle\langle u_{j}|-\mathrm{H.c}\right].\label{eq:tot_Ham}
\end{align}

Writing the time-dependent state vector as $\left|\psi(t)\right\rangle =\sum_{k}c_{k}(t)|u_{k}\rangle$,
we arrive at the Schr\"{o}dinger equation of the coefficients $\{c_{k}\}$:
\begin{equation}
\frac{d}{dt}c_{k}(t)=-iE_{k}c_{k}(t)-\varepsilon_{D}\sum_{j}\left[e^{i\omega_{D}t}\gamma_{kj}-\mathrm{H.c.}\right]c_{j}(t).\label{eq:Sch_eqn}
\end{equation}
In the following, the determination of entanglement will be carried
out from the state coefficients under the bare-state basis, i.e. transforming
back the dressed states, we have
\begin{equation}
\beta_{m,n}(t)=\sum_{k}c_{k}(t)\alpha_{m,n}^{(k)}\label{eq:system_state}
\end{equation}
for the vector $|\psi(t)\rangle=\sum_{\left\langle m\right\rangle }\beta_{m,0}(t)\left|\phi_{m},0\right\rangle +\beta_{m,1}(t)\left|\phi_{m},1\right\rangle $.

\section{Evolution and four-tangle\label{sec:evolution}}

The partitioning of the bipartite and the quadripartite entanglements
that evolve with time is reflected in the coefficients $\beta_{m,n}(t)$.
To be exact, we follow the definition of polynomial invariant~\citep{Eltschka12}
that generalizes 3-tangle~\citep{coffman00} to measure distirbuted
entanglement in arbitrary $N$-partite systems. Here we customize
this polynomial invariant to degree 4 to have
\begin{equation}
|\mathcal{H}(t)|{}^{2}=\frac{1}{2}\sum_{j=1}^{3}(-1)^{j+1}\sum_{\{j\}}C_{\{j\}|\{4-j\}}^{2}\left[\psi(t)\right]\label{eq:invar}
\end{equation}
to reflect the entanglement distribution, thus the degree of monogamy,
in the quadripartite system under study. In this customization,
\begin{equation}
C_{\{j\}|\{4-j\}}^{2}\left[\psi\right]=2\left(1-\mathrm{tr}\rho_{\{j\}}^{2}\right)
\end{equation}
indicates the concurrence between a $j$-component subsystem and the
rest parts~\citep{coffman00,Eltschka12}, where $\rho_{\{j\}}$ denotes
the reduced density matrix for the $j$ components with the rest $(4-j)$
components traced out. The sum over $\{j\}$ in Eq.~(\ref{eq:invar})
is taken over all combinations of $j$ components out of the four
(e.g. when $j=2$, the index \{2\} includes the combination of qubit
$L$ and qubit $R$). To simplify the terminology, we shall call Eq.~(\ref{eq:invar})
4-tangle in the discussion below.

In the quadripartite system, the cavity field in the stripline resonator
acts as a quantum bus that simultaneously couples to all three qubits.
It serves, therefore, as a mediator that distributes entanglement
among all components it couples to, similar to the role played by
the mechanical resonator in a double-optical-cavity system~\citep{Huan15}.
Here, being driven by an external microwave field from the waveguide,
the cavity field has its state vary over time and hence redistributes
the entanglement among the qubits over time.

\begin{figure}
\includegraphics[bb=8bp 10bp 660bp 500bp,clip,width=8.2cm]{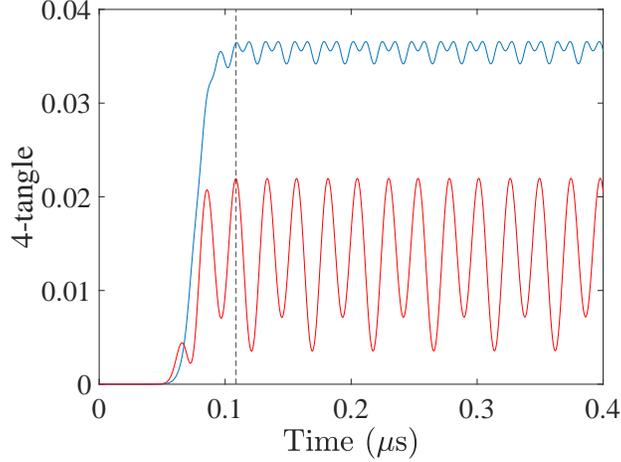}\caption{4-tangle initiated from a populated one-photon state at the cavity
field, $\left|\psi(0)\right\rangle =\left|g,g,g,1\right\rangle $,
which shows a finite duration of synchronization (up to the dashed
line) before the entanglement reaches a saturated value. The saturation
signifies the completion of synchronization, which are visible for
both a homogeneous coupling scenario (blue curve) and an inhomogeneous
coupling scenario (red curve). The parameters used in generating the
plot are given in the text.~\label{fig:initiation}}
\end{figure}

To observe the process of the redistribution and to decide whether
the monogamous relation is obeyed, we consider an initial state which
has the cavity mostly populated to initiate the entanglement. With
full population at the one-photon state, i.e. $\left|\psi(0)\right\rangle =\left|\phi_{0},1\right\rangle $,
the distributed entanglement measured by the 4-tangle, as shown in
Fig.~\ref{fig:initiation}, reaches a saturated value after the cavity
field synchronizes the evolution of the qubits~\citep{Huan20}. The
saturated value is defined as the periodic peak obtainable by the
4-tangle and the synchronization duration is then the time between
the start of the entanglement and the moment at which the first one
of such peaks appears. Besides the rising time and the saturated value,
the synchronization pattern is typical whether the couplings among
the qubits are homogeneous or inhomogeneous. In the plot, we used
experimentally accessible transition frequencies of transmon qubits
at $\Omega_{L}/2\pi=\Omega_{R}/2\pi=6.112$~GHz and, to account for
the discrepancies at fabrication~\citep{fink09}, have let $\Omega_{M}/2\pi=6.111$~GHz.
The cavity is slightly detuned from the qubits at $\omega_{c}/2\pi=6.13$~GHz.
The microwave field in the waveguide drives the cavity at $\varepsilon_{D}/2\pi=200$~kHz
and propagates at $\omega_{D}/2\pi=6.11$~GHz. For the homogeneous
case, the coupling strength is set to $\eta_{\nu}/2\pi=300$~MHz
for all $\nu$ among $\{L,M,R\}$; for the inhomogeneous case, $\eta_{L}/2\pi=\eta_{R}/2\pi=300$~MHz
and $\eta_{M}/2\pi=150$~MHz. The differing aspects in the two scenarios
is that homogeneous coupling permits a greater saturated 4-tangle
at the expense of a slower rising time.

\section{Periodicity in monogamy\label{sec:periodicity}}

From Fig.~\ref{fig:initiation}, we also observe periodicity in the
variation, akin to the vanishing and revival effects observed in other
entanglement studies, albeit neither case has the entanglement measure
completely vanish where the monogamy relation would reduce to its
equality limit. We find that the evolution of the 4-tangle in the
quadripartite system is highly dependent on the initial state. With
a slight alteration to the cavity photon, by letting $\beta_{0,1}=\sqrt{0.8}$
while having qubit $L$ slightly inverted with $\beta_{4,0}=\beta_{4,1}=\sqrt{0.1}$
at initial time, the monogamy equality can be asymptotically achieved,
where the periodicity depends on all three coupling strengths $\eta_{L}$,
$\eta_{R}$, and $\eta_{M}$.

\begin{figure}
\includegraphics[bb=0bp 0bp 590bp 480bp,clip,width=8.1cm]{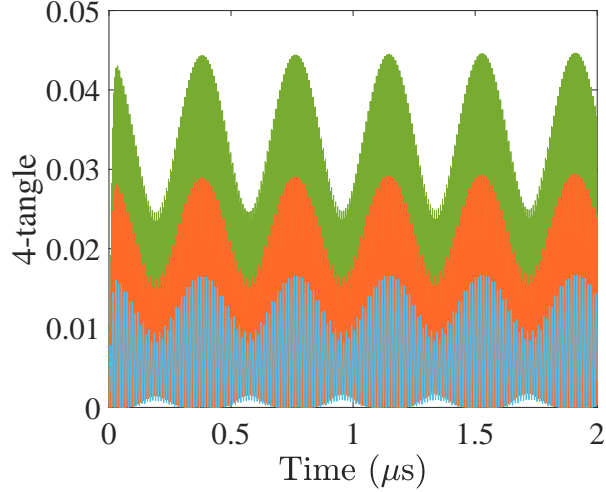}\caption{Time evolutions of the 4-tangle for three different homogeneous ($\lambda=1$)
coupling strengths: $\eta_{M}/2\pi=300$~MHz (blue curve), $400$~MHz
(orange curve), and $500$~MHz (green curve). When all three qubits
stay at the adjacent antinodes of the cavity field, the period of
the 4-tangle is not affected by the magnitude of $\eta_{M}$ but the
peak value of the 4-tangle increases with $\eta_{M}$.~\label{fig:same lambda}}
\end{figure}

To obtain an appropriate coupling combination for a desired pair of
revival period and 4-tangle magnitude, one should scan over the triple
parameter space ($\eta_{L},\eta_{R},\eta_{M}$). Nevertheless, since
our goal is to seek the effect of asynchronous excitation on entanglement
generation, the parameter space is compressed to 2-dimensional $(\lambda,\eta_{L})$
to simplify our study. We let qubits $L$ and $R$ be fixated at antinodes
to receive the same maximal coupling ($\eta_{L}=\eta_{R}$) while
allowing qubit $M$ to be removed from antinode to receive sub-maximal
coupling characterized by the dimensionless parameter $\lambda=\eta_{M}/\eta_{L}$.
In other words, the excitation rate of qubit $M$ is asynchronous
with qubits $L$ and $M$ where the pair $(\lambda,\eta_{L})$ signifies
the absolute coupling strength and the inhomogeneity of the couplings.

We first consider the homogeneous coupling scenario, i.e. $\lambda=1$.
As shown in Fig.~\ref{fig:same lambda}, the time evolution of the
4-tangle follows the pattern of Fig.~\ref{fig:initiation} for all
$\eta_{M}$, which demonstrate periodic vanishing and revival patterns
after a short duration of rising from initial zero value. Throughout,
the alternating-sign sum of concurrences is found to be always non-negative,
preserving the monogamy inequality $|\mathcal{H}(t)|{}^{2}>0$ in
Eq.~(\ref{eq:invar}). Furthermore, irrespective of the coupling
strength, the monogamy equality limit $|\mathcal{H}(t)|{}^{2}=0$
is reached at the same time instants. With the same system parameters
as in Fig.~\ref{fig:initiation}, the period $\tau$ is measured
at $0.348\thinspace\mu$s. The amplitude of 4-tangle $|\mathcal{H}(t)|{}^{2}$
monotonically follows the coupling strength $\eta_{M}$.

\begin{figure}
\includegraphics[bb=15bp 5bp 600bp 505bp,clip,width=8.1cm]{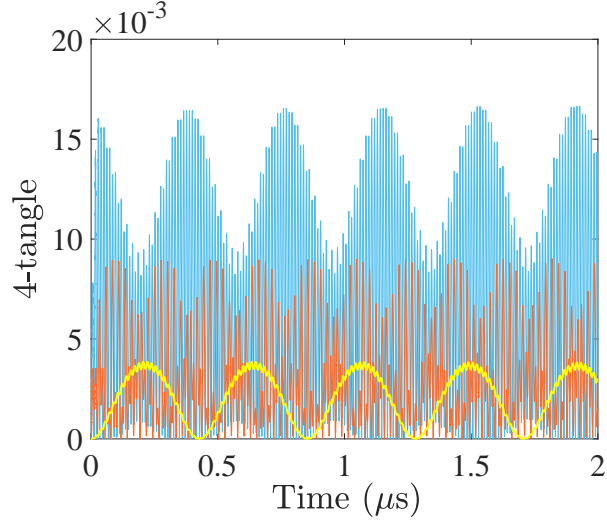}\caption{Time evolutions of 4-tangle for three different cases $\lambda=0.05$
(yellow curve), $\lambda=0.5$ (orange curve), and $\lambda=1$ (blue
curve) of homogeneity in coupling where $\eta_{L}/2\pi=\eta_{R}/2\pi=300$~MHz.
The periodicity in death and revival of entanglement is prevalent
for both the homogeneous coupling ($\lambda=1$) and the inhomogeneous
couplings ($\lambda=0.5$ and $0.05$). Out of the three cases shown,
the period $\tau$ for $\lambda=0.05$ is the largest and that for
$\lambda=0.5$ is the smallest, showing a nonlinear relation between
$\tau$ and $\lambda$.~\label{fig:same eta}}
\end{figure}

For the inhomogeneous scenario, which can be implemented by removing
the qubit $M$ from the antinode of the cavity field as indicated
in Fig.~\ref{fig:model}, the periodic patterns of the 4-tangle evolution
are reflected in Fig.~\ref{fig:same eta}. In the plot, the coupling
strengths $\eta_{L}/2\pi$ and $\eta_{R}/2\pi$ are let fixed at $300$~MHz,
while the relative coupling parameter $\lambda$ takes the values
$0.05$, $0.5$, and $1$. The unity case (given by the blue curve)
indicates the homogeneous coupling and is the same of the one shown
in Fig.~\ref{fig:same lambda}. Using it as a reference, we observe
that lowering $\lambda$ and thereby permitting inhomogeneous excitation
to the middle qubit leads to a monotonic decrease in the oscillating
amplitude of the entanglement, but the period of oscillation is affected
in a non-monotonic way.

During the evolution, the amplitude of oscillation in the 4-tangle
varies over time while the period between one asymptotic vanishing
and the next remains fixed. When $\lambda$ is reduced from 1 to 0.5,
the maximum amplitude decreases to about half the original amplitude
while the period $\tau$ is reduced from $0.382\thinspace\mu$s to
$0.193\thinspace\mu$s. When $\lambda$ is further reduced from 0.5
to 0.05, the maximum amplitude is reduced by about 95\%, whereas the
period $\tau$, on the contrary, increases from $0.193\thinspace\mu$s
to $0.428\thinspace\mu$s.

\begin{figure}
\includegraphics[bb=16bp 10bp 400bp 550bp,clip,width=8.5cm]{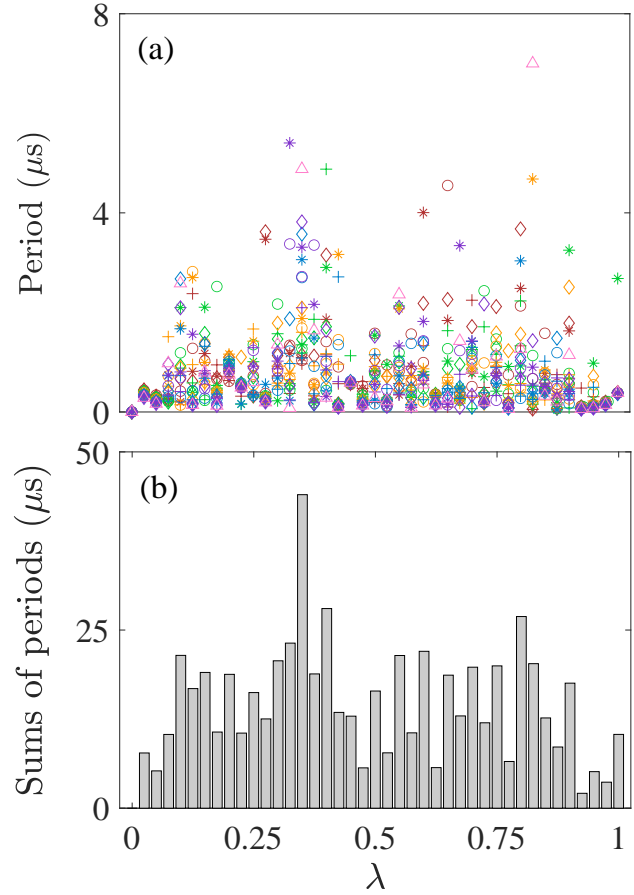}\caption{(a) Period of the 4-tangle variation as a function of the relative
coupling $\lambda$. Each data point is obtained from a simulation
conducted as in Fig.~\ref{fig:same eta} by varying $\lambda$ at
different coupling strength $\eta_{L}/2\pi$, ranging from $300$~MHz
to $500$~MHz, at $10$~MHz intervals. The data points are differentiated
by both color and symbols used: those between $300$~MHz to $330$~MHz
are colored red, between $340$~MHz and $370$~MHz green, between
$380$~MHz and $410$~MHz yellow, between $420$~MHz and $450$~MHz
blue, and between $460$~MHz and $490$~MHz purple. Those of $500$~MHz
are colored pink. The symbols within each band follow the order $\{+,\bigcirc,*,\diamondsuit\}$,
from small to large. (b) Histogram of the slotted or binned values
of $\lambda$, where the periods within each column shown in (a) are
summed into separate slots.~\label{fig:bar}}
\end{figure}

Overall, the dependence of the period $\tau$ on the $(\lambda,\eta_{L})$
is highly nonlinear and not extractable analytically from the expression
of Eq.~(\ref{eq:invar}). We resort to a statistical method to characterize
this dependence. We have computed the evolutions of the 4-tangle when
$\lambda$ varies between zero and one over a range of values of coupling
strength $\eta_{L}$ and extracted the periods from the plots for
different combinations of $\lambda$ and $\eta_{L}$. Using a scatter
plot of Fig.~\ref{fig:bar}, we mark each extracted period as a data
point, which is color- and symbol-coded as in Fig.~\ref{fig:bar}(a),
and binned the data points into slots each differing from the neighboring
slot by $0.025$, within which the values of the data points are summed
as in Fig.~\ref{fig:bar}(b). For each slot of $\lambda$, $\eta_{L}/2\pi$
varies between $300$~MHz and $500$~MHz at $10$~MHz intervals
and the other system parameters remain identical to those used in
the figures above.

Therefore, Fig.~\ref{fig:bar} shows on average how likely a combination
of $(\lambda,\eta_{L})$ would generate a longer period in 4-tangle.
We observe that, statistically speaking, this period maximizes at
$\lambda\approx0.35$, where the scatter datapoints have the greatest
accumulated value and thus one is most likely to obtain a long revival
period at this $\lambda$ regardless the absolute coupling strength
$\eta_{L}$. In constrast, it is less likely at $\lambda=0.5$ and
least likely at $\lambda\approx0.925$. In particular, though one
datapoint at $\lambda\approx0.8$ corresponds to a long period in
(a), its binned sum is less than that of $\lambda\approx0.35$, showing
it is less likely on average to obtain a long period at $\lambda\approx0.8$
when all $\eta_{L}$ values are considered.

\begin{figure}
\includegraphics[bb=20bp 0bp 750bp 530bp,clip,width=8.8cm]{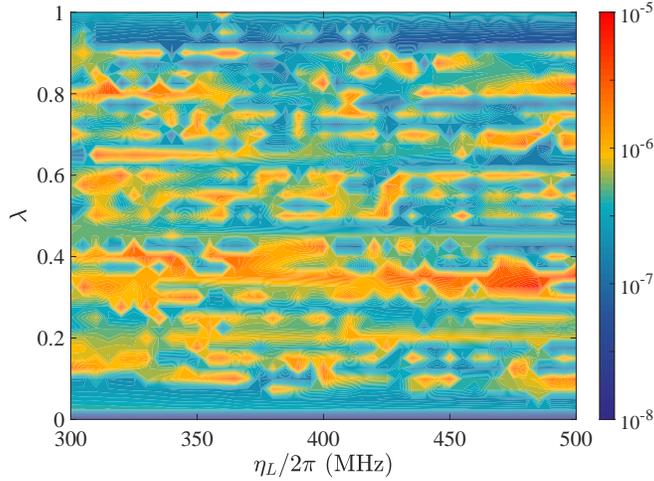}\caption{The filled contour map measuring the 4-tangle on a log-scaled color
axis against the absolute coupling strength $\eta_{L}$ on the horizontal
axis and the relative coupling $\lambda$ on the vertical axis.~\label{fig:contour}}
\end{figure}

To obtain a better resolution of the variation of period $\tau$ against
$\eta_{L}$, we have further conducted simulations running at $5$~MHz
intervals for $\eta_{L}$, while retaining an interval of $0.025$
along the $\lambda$-axis, and summarize the results in the contour
plot in Fig.~\ref{fig:contour}. The period $\tau$ is log-scaled
and plotted in color against $\lambda$ and $\eta_{L}$. First, verifying
the findings from Fig.~\ref{fig:bar}, $\tau$ maximizes at the inhomogeneous
couplings of $\lambda\approx0.35$ while minimizes at the two ends
and the middleway of $\lambda$. Secondly, at the maximizing values
of $\lambda$, the dependence of $\tau$ on the absolute coupling
$\eta_{L}$ is not uniform, showing also a nonlinear relationship.
Therefore, if one considers prolonging the duration of entanglement
for the purpose of processing quantum information in a multi-qubit
system, uniform couplings among the qubits are not necessarily beneficial.
Rather, inhomogeneous coupling peculiar to the system setting can
provide a means of assistance. For example, for the quadripartite
qubit-cavity system under study, given the same transmon qubit transition
frequencies as in Sec.~\ref{sec:evolution}A, the longest period
$\tau$ appears at $\eta_{L}/2\pi=485$~MHz. For a stripline cavity
of length $24$~mm~\citep{wallraff04}, this signifies a setup that
prolongs the duration of entanglement by moving the middle qubit as
shown in Fig.~\ref{fig:model} to the position $4.53$~mm from the
center antinode.

\section{Conclusions\label{sec:Conclusions}}

We have studied the dynamic evolution of the entanglement measure
4-tangle, which is a polynomial invariant that quantifies the degree
of separation of the monogamous entanglements from the group entanglements,
for a quadripartite system consisting of three qubits and one cavity
mode. We find that the 4-tangle throughout the interactive evolution
of the qubits is guaranteed non-negative and exhibits periodic revival.
Under the framework of circuit QED with three qubits, it is shown
that it inhomogeneity of couplings that induces asynchronous excitations
to the qubits facilitate the choice of entanglement generation. By
selecting a combination of absolute relative coupling strengths among
the qubits, one can not only obtain the desired distributed entanglement
for generating GHZ states, but also determine how soon the entanglement
is realized.

For instance, in some cases when fast generation of entanglement is
desired, one can place all three qubits on the antinodes for homogeneous
maximal couplings or allow the middle qubit to be placed midway towards
a neighboring node where the effective coupling is half of the possible
maximum. In some other cases, slow generation might be wanted, such
as when longer rise and fall times in entanglement are desired to
tolerate the timing inaccuracies of the experimental apparatus, so
that the fidelity of actual target state would be improved. In such
cases, one can move the middle qubit to the place where the coupling
is about one-third of the maximum while retaining the other two qubits
at the antinodes.

We note that the study here is limited to three cavity-coupled qubits.
Generalization to larger number of qubits with a parameter space of
higher dimensions require future works in this direction. Also, the
methodology we use here is based on statistical and numerical analysis.
Alternative approaches are needed to seek an analytical optimization
method for finding entanglement characteristics such as the extremal
periods.
\begin{acknowledgments}
H. I. thanks the support by the Science and Technology Development
Fund, Macau SAR (File no. 0130/2019/A3) and by University of Macau
(MYRG2018-00088-IAPME).
\end{acknowledgments}


\begin{thebibliography}{99}
\bibitem{Chen07}K. Chen and H.-K. Lo, Quan. Inf. Comput. \textbf{7},
689 (2007).

\bibitem{Gao05}T. Gao, F. L. Yan, and Z. X. Wang, J. Phys. A: Math.
Gen. \textbf{38}, 5761 (2005).

\bibitem{chin12}A. W. Chin, S. F. Huelga, and M. B. Plenio, Phys.
Rev. Lett. \textbf{109}, 233601 (2012). 

\bibitem{coffman00} V. Coffman, J. Kundu, and W. K. Wootters, Phys.
Rev. A \textbf{61}, 052306 (2000).

\bibitem{Wootters98}W. K. Wootters, Phys. Rev. Lett. \textbf{80},
2245 (1998).

\bibitem{Horodecki09}R. Horodecki, P. Horodecki, M. Horodecki, and
K. Horodecki, Rev. Mod. Phys. \textbf{81}, 865 (2009).

\bibitem{Osborne06} T. J. Osborne and F. Verstraete, Phys. Rev. Lett.
\textbf{96}, 220503 (2006).

\bibitem{Cornelio13} M. F. Cornelio, Phys. Rev. A \textbf{87},032330
(2013).

\bibitem{Eltschka12}C. Eltschka, T. Bastin, A. Osterloh, and J. Siewert,
Phys. Rev. A \textbf{85}, 022301 (2012). 

\bibitem{Wong01}A. Wong and N. Christensen, Phys. Rev. A \textbf{63},
044301 (2001). 

\bibitem{Mintert051} F. Mintert, M. Kuś, and A. Buchleitner, Phys.
Rev. Lett. \textbf{95}, 260502 (2005).

\bibitem{Mintert052} F. Mintert, A. R. R. Carvalho, M. Kuś, and A.
Buchleitner, Phys. Rep. \textbf{415}, 207 (2005).

\bibitem{Dur00}W. Dür, G. Vidal, and J. I. Cirac, Phys. Rev. A \textbf{62},
062314 (2000).

\bibitem{Regula16}B. Regula, A. Osterloh, and G. Adesso, Phys. Rev.
A \textbf{93}, 052338 (2016).

\bibitem{Gartzke18}S. Gartzke and A. Osterloh, Phys. Rev. A \textbf{98},
052307 (2018).

\bibitem{Oliveira14} T. R. de Oliveira, M. F. Cornelio, and F. F.
Fanchini, Phys. Rev. A \textbf{89}, 034303 (2014).

\bibitem{Bai14} Y.-K. Bai, Y.-F. Xu, and Z. D. Wang, Phys. Rev. Lett.
\textbf{113}, 100503 (2014).

\bibitem{Liu17}Y. Liu, S. Kuang, and S. Cong, IEEE Trans. Cyber.
\textbf{47}, 3827 (2017). 

\bibitem{Huan15}T. Huan, R. Zhou, and H. Ian, Phys. Rev. A \textbf{92},
022301 (2015).

\bibitem{Huan20}T. Huan, R. Zhou, and H. Ian, Sci. Rep. \textbf{10},
12975 (2020).

\bibitem{Nourman16}A. Nourmandipour, M. K. Tavassoly, and M. Rafiee,
Phys. Rev. A 93, 022327 (2016).

\bibitem{Zhang10} Y.-J. Zhang, Z.-X. Man, X.-B. Zou, Y.-J. Xia, and
G.-C. Guo, J. Phys. B: At. Mol. Opt. Phys. \textbf{43}, 045502 (2010). 

\bibitem{wallraff04}A. Wallraff, D. I. Schuster, A. Blais, L. Frunzio,
R.-S. Huang, J. Majer, S. Kumar, S. M. Girvin, and R. J. Schoelkopf,
Nature \textbf{431}, 162 (2004).

\bibitem{fink09}J. M. Fink, R. Bianchetti, M. Baur, M. Göppl, L.
Steffen, S. Filipp, P. J. Leek, A. Blais, and A. Wallraff, Phys. Rev.
Lett. \textbf{103}, 083601 (2009).

\bibitem{ian12}H. Ian, Y. Liu, and F. Nori, Phys. Rev. A \textbf{85},
053833 (2012).

\bibitem{ian14}H. Ian and Y. Liu, Phys. Rev. A \textbf{89}, 043804
(2014).

\bibitem{Ian16}H. Ian, EPL \textbf{114}, 50005 (2016).

\bibitem{ficek06}Z. Ficek and R. Tanaś, Phys. Rev. A \textbf{74},
024304 (2006).

\bibitem{Dukalski10}M. Dukalski and Ya. M. Blanter, Phys. Rev. A
\textbf{82}, 052330 (2010).

\bibitem{Wang14}G. Wang, L. Huang, Y.-C. Lai, and C. Grebogi, Phys.
Rev. Lett. \textbf{112}, 110406 (2014).

\bibitem{kjaergaard20}M. Kjaergaard, M. E. Schwartz, J. Braumüller,
P. Krantz, J. I.-J. Wang, S. Gustavsson, and W. D. Oliver, Superconducting
Qubits: Current State of Play, Annu. Rev. Condens. Matter Phys. \textbf{11},
369 (2020).

\bibitem{Borkje13}K. Børkje, A. Nunnenkamp, J. D. Teufel, and S.
M. Girvin, Phys. Rev. Lett. \textbf{111}, 053603 (2013).
\end{thebibliography}
\end{document}